\input harvmac
\overfullrule=0pt

\def\frac#1#2{{#1\over #2}}

\def\OO{{\cal O}}
\def\ie{{\it i.e.}}
\def\eg{{\it e.g.}}
\def\CM{{\cal M}}



\Title{\vbox{\baselineskip12pt \hbox{hep-th/0312098}
\hbox{EFI-03-46}	
}}
{\vbox{
\centerline{New Results on the ``$a$-theorem'' in}
\vskip 10pt
\centerline{Four Dimensional 
Supersymmetric Field Theory}
}}
\smallskip
\centerline{D. Kutasov}
\medskip

\centerline{\it  Enrico Fermi Institute and Dept. of Physics, University of
Chicago} 
\centerline{\it  5640 S. Ellis Av., Chicago, IL 60637, USA}

\bigskip
\noindent
In four dimensional $N=1$ supersymmetric field theory it is often the case
that the $U(1)_R$ current that becomes part of the superconformal 
algebra at the infrared fixed point is conserved throughout the renormalization
group (RG) flow. We show that when that happens, the central charge $a$  
decreases under RG flow. The main tool we employ is an extension of recent 
ideas on ``$a$-maximization'' away from fixed points of the RG. This extension
is useful more generally in studying RG flows in supersymmetric theories.

\Date{December 2003}

\vfil\eject


\lref\ZamolodchikovGT{
A.~B.~Zamolodchikov,
``'Irreversibility' Of The Flux Of The Renormalization Group In A 2-D Field Theory,''
JETP Lett.\  {\bf 43}, 730 (1986)
[Pisma Zh.\ Eksp.\ Teor.\ Fiz.\  {\bf 43}, 565 (1986)].
}

\lref\KutasovSV{
D.~Kutasov and N.~Seiberg,
``Number Of Degrees Of Freedom, Density Of States And Tachyons In String Theory And Cft,''
Nucl.\ Phys.\ B {\bf 358}, 600 (1991).
}

\lref\CardyCW{
J.~L.~Cardy,
``Is There A C Theorem In Four-Dimensions?,''
Phys.\ Lett.\ B {\bf 215}, 749 (1988).
}

\lref\JackEB{
I.~Jack and H.~Osborn,
``Analogs For The C Theorem For Four-Dimensional Renormalizable Field Theories,''
Nucl.\ Phys.\ B {\bf 343}, 647 (1990).
}

\lref\CappelliYC{
A.~Cappelli, D.~Friedan and J.~I.~Latorre,
``C Theorem And Spectral Representation,''
Nucl.\ Phys.\ B {\bf 352}, 616 (1991).
}

\lref\CappelliKE{
A.~Cappelli, J.~I.~Latorre and X.~Vilasis-Cardona,
``Renormalization group patterns and C theorem in more than two-dimensions,''
Nucl.\ Phys.\ B {\bf 376}, 510 (1992)
[arXiv:hep-th/9109041].
}

\lref\BastianelliVV{
F.~Bastianelli,
``Tests for C-theorems in 4D,''
Phys.\ Lett.\ B {\bf 369}, 249 (1996)
[arXiv:hep-th/9511065].
}

\lref\ForteDX{
S.~Forte and J.~I.~Latorre,
``A proof of the irreversibility of renormalization group flows in four  dimensions,''
Nucl.\ Phys.\ B {\bf 535}, 709 (1998)
[arXiv:hep-th/9805015].
}

\lref\AnselmiYS{
D.~Anselmi, J.~Erlich, D.~Z.~Freedman and A.~A.~Johansen,
``Positivity constraints on anomalies in supersymmetric gauge theories,''
Phys.\ Rev.\ D {\bf 57}, 7570 (1998)
[arXiv:hep-th/9711035].
}

\lref\AnselmiAM{
D.~Anselmi, D.~Z.~Freedman, M.~T.~Grisaru and A.~A.~Johansen,
``Nonperturbative formulas for central functions of supersymmetric gauge  theories,''
Nucl.\ Phys.\ B {\bf 526}, 543 (1998)
[arXiv:hep-th/9708042].
}

\lref\AnselmiUK{
D.~Anselmi,
``Quantum irreversibility in arbitrary dimension,''
Nucl.\ Phys.\ B {\bf 567}, 331 (2000)
[arXiv:hep-th/9905005].
}

\lref\CappelliDV{
A.~Cappelli, G.~D'Appollonio, R.~Guida and N.~Magnoli,
``On the c-theorem in more than two dimensions,''
arXiv:hep-th/0009119.
}

\lref\SeibergPQ{
N.~Seiberg,
``Electric - magnetic duality in supersymmetric nonAbelian gauge theories,''
Nucl.\ Phys.\ B {\bf 435}, 129 (1995)
[arXiv:hep-th/9411149].
}

\lref\IntriligatorJJ{
K.~Intriligator and B.~Wecht,
``The exact superconformal R-symmetry maximizes a,''
Nucl.\ Phys.\ B {\bf 667}, 183 (2003)
[arXiv:hep-th/0304128].
}

\lref\KutasovIY{
D.~Kutasov, A.~Parnachev and D.~A.~Sahakyan,
``Central charges and U(1)R symmetries in N = 1 super Yang-Mills,''
JHEP {\bf 0311}, 013 (2003)
[arXiv:hep-th/0308071].
}

\lref\IntriligatorMI{
K.~Intriligator and B.~Wecht,
``RG fixed points and flows in SQCD with adjoints,''
arXiv:hep-th/0309201.
}

\lref\KutasovVE{
D.~Kutasov,
``A Comment on duality in N=1 supersymmetric nonAbelian gauge theories,''
Phys.\ Lett.\ B {\bf 351}, 230 (1995)
[arXiv:hep-th/9503086].
}

\lref\KutasovNP{
D.~Kutasov and A.~Schwimmer,
``On duality in supersymmetric Yang-Mills theory,''
Phys.\ Lett.\ B {\bf 354}, 315 (1995)
[arXiv:hep-th/9505004].
}

\lref\KutasovSS{
D.~Kutasov, A.~Schwimmer and N.~Seiberg,
``Chiral Rings, Singularity Theory and Electric-Magnetic Duality,''
Nucl.\ Phys.\ B {\bf 459}, 455 (1996)
[arXiv:hep-th/9510222].
}

\lref\BrodieVX{
J.~H.~Brodie,
``Duality in supersymmetric SU(N/c) gauge theory with two adjoint chiral  superfields,''
Nucl.\ Phys.\ B {\bf 478}, 123 (1996)
[arXiv:hep-th/9605232].
}

\lref\NovikovUC{
V.~A.~Novikov, M.~A.~Shifman, A.~I.~Vainshtein and V.~I.~Zakharov,
``Exact Gell-Mann-Low Function Of Supersymmetric Yang-Mills Theories From Instanton Calculus,''
Nucl.\ Phys.\ B {\bf 229}, 381 (1983).
}

\lref\ShifmanZI{
M.~A.~Shifman and A.~I.~Vainshtein,
``Solution Of The Anomaly Puzzle In Susy Gauge Theories And The Wilson Operator Expansion,''
Nucl.\ Phys.\ B {\bf 277}, 456 (1986)
[Sov.\ Phys.\ JETP {\bf 64}, 428 (1986\ ZETFA,91,723-744.1986)].
}

\lref\GrossCS{
D.~J.~Gross and F.~Wilczek,
``Asymptotically Free Gauge Theories. 2,''
Phys.\ Rev.\ D {\bf 9}, 980 (1974).
}

\lref\BanksNN{
T.~Banks and A.~Zaks,
``On The Phase Structure Of Vector - Like Gauge Theories With Massless Fermions,''
Nucl.\ Phys.\ B {\bf 196}, 189 (1982).
}

\lref\JackCN{
I.~Jack, D.~R.~T.~Jones and C.~G.~North,
``Scheme dependence and the NSVZ beta-function,''
Nucl.\ Phys.\ B {\bf 486}, 479 (1997)
[arXiv:hep-ph/9609325].
}

\lref\LeighEP{
R.~G.~Leigh and M.~J.~Strassler,
``Exactly marginal operators and duality in 
four-dimensional N=1 supersymmetric gauge theory,''
Nucl.\ Phys.\ B {\bf 447}, 95 (1995)
[arXiv:hep-th/9503121].
}

\newsec{Introduction}

A Quantum Field Theory (QFT) is traditionally defined as an ultraviolet (UV)
fixed point connected to an infrared (IR) fixed point by an RG flow. At the UV
fixed point, which describes the very short distance physics, the theory is 
scale invariant, and in all known examples conformal. A non-trivial RG flow
is typically induced by perturbing the UV fixed point by adding to the
Lagrangian a relevant\foot{Here and below ``relevant'' includes ``marginally
relevant,'' as in asymptotically free gauge theories.} operator. In the presence of
the perturbation, the short distance behavior of correlation functions is still
described by the UV conformal field theory (CFT), while at long distances the
theory flows to an IR fixed point, where it becomes conformal again.

RG flow is in general irreversible. For example, if CFT $B$ is 
obtained by perturbing CFT $A$ by a relevant operator and going 
to long distances, one cannot (in all known examples) get back from
$B$ to $A$ by perturbing $B$ by a relevant operator.  Therefore, 
it is natural to ask whether one can define an intrinsic characteristic 
of fixed points which keeps track of this irreversibility -- a real 
number $\CM$ associated with each CFT which has the property that if 
CFT's $A$ and $B$ are connected by an RG flow and $\CM(A)>\CM(B)$, 
it must be that $A$ is the UV fixed point of the flow and $B$ the IR  
one, and vice versa. It would be nice if such an $\CM$ also counted 
the ``number of degrees of freedom'' of the fixed point, due to the 
intuition that RG flow proceeds by decoupling of high energy degrees 
of freedom as the distance scale is increased. A necessary condition 
for such an interpretation seems to be that $\CM(A)$ should be positive 
for all\foot{We restrict attention to unitary theories.} CFT's $A$.

In two dimensional QFT it was shown in \ZamolodchikovGT\ that one can 
choose $\CM$ to be the Virasoro central charge $c$. The Zamolodchikov 
$c$-theorem states that  one can define a quantity with the following 
properties:
\item{(1)}  It is positive and monotonically decreasing throughout the RG flow.
\item{(2)}  At the UV and IR fixed points it coincides with the central charge
of the relevant CFT, $c$.

\noindent
Moreover, at fixed points of the RG one can think of the central charge as 
counting the number of degrees of freedom of the theory \KutasovSV.

Despite much work (see \eg\ 
\refs{\CardyCW\JackEB\CappelliYC\CappelliKE\BastianelliVV\AnselmiAM
\AnselmiYS\ForteDX\AnselmiUK-\CappelliDV}), an analog of the $c$-theorem
in four dimensional QFT has not been proven so far. At the same time, increasing
evidence has been accumulating for the conjecture \CardyCW, often refered to as
the ``$a$-theorem,'' that the analog of $c$ in four dimensions is the
central charge $a$, the coefficient of the Euler density in the 
conformal anomaly on a curved spacetime manifold. By studying examples
of RG flows for which the UV and IR fixed points are sufficiently well 
understood to compute the value of $a$, it was found that in all cases 
$a_{UV}>a_{IR}$.

Most of the examples alluded to above correspond to supersymmetric 
field theories. One reason for that is that our understanding of 
four dimensional supersymmetric field theories is significantly 
better than that of non-supersymmetric ones, and in many cases we 
now understand the infrared behavior of these theories well enough 
to compute $a$. Another reason is that in supersymmetric theories 
the conformal anomaly that gives rise to $a$  is related by 
supersymmetry to an anomaly associated with the $U(1)_R$ current 
that belongs to the $N=1$ superconformal multiplet 
\refs{\AnselmiAM\AnselmiYS}. Therefore, if we can identify the 
superconformal $U(1)_R$ current at the UV and IR fixed points of a 
supersymmetric RG flow, we can check whether the $a$-theorem is satisfied. 

There are two basic tools that have been used to determine the superconformal 
$U(1)_R$ symmetry at strongly coupled fixed points. One is Seiberg duality 
\SeibergPQ, which relates the infrared behavior of different gauge theories. 
This duality often relates a theory that is strongly coupled in some variables 
to a theory which is weakly coupled, or free, in other variables.

The second tool, which is more important for the purpose of the present 
discussion, is the following. It has been known for a long time that in 
many interacting four dimensional QFT's, the $U(1)_R$ that becomes part 
of the superconformal group at the IR fixed point is conserved throughout 
the RG flow, \ie\ it is part of the symmetry group of the full theory. 
The symmetry of the full theory can be analyzed by studying the vicinity 
of the UV fixed point, where the dynamics is often simpler due to asymptotic 
freedom. If the symmetry group contains a unique $U(1)_R$ which satisfies
the physical consistency conditions, one can identify it with the IR 
superconformal R-symmetry, and use `t Hooft anomaly matching to compute
the central charge $a$. 

A major stumbling block in implementing the second method of 
determining the superconformal R symmetry has been that in
many cases there is more than one $U(1)_R$ that is preserved 
throughout the RG flow and satisfies all the necessary conditions. 
Progress on this problem was recently reported by Intriligator and Wecht
(IW), who showed that there are in fact additional consistency conditions, 
that were not taken into account previously, which uniquely determine the 
superconformal R-charge in these cases \IntriligatorJJ.  Let $R$ be any of 
the global R-charges which are conserved throughout the RG flow. Compute 
the central charge $a$, which is given\foot{After rescaling by a factor
$32/3$.} by a particular combination of `t Hooft anomalies
\eqn\aaa{a=3 {\rm tr} R^3-{\rm tr}R~.}
The fact that there is more than one possible choice of $U(1)_R$ implies that 
the anomaly $a$ depends on some continuous parameters, which parametrize the 
particular $U(1)_R$ whose anomaly is being computed.  IW proved that the 
superconformal $U(1)_R$ of the infrared theory is the unique current that 
corresponds to a local maximum of $a$ \aaa\ as a function of all the continuous 
parameters mentioned above.

The results of IW allow one to obtain a more detailed understanding of the 
infrared behavior and phase structure of many gauge theories that 
were previously mysterious. Recent investigations using these results 
\refs{\KutasovIY,\IntriligatorMI} revealed a rich structure of fixed points
and flows between them, consistent with previous work on Seiberg duality 
\refs{\SeibergPQ,\KutasovVE\KutasovNP\KutasovSS-\BrodieVX} and with
the predictions of the $a$-theorem.

A natural question that arises from the results of
\refs{\IntriligatorJJ-\IntriligatorMI} is
whether they imply the $a$-theorem, at least within
their domain of validity. In this note we will show
that they do. More precisely, we will prove that if
the $U(1)_R$ that becomes part of the superconformal
algebra in the IR is preserved throughout the RG flow, then
\eqn\athm{a_{UV}>a_{IR}~.}
The main observation we will use is that the idea of
$a$-maximization introduced in \IntriligatorJJ\ can
be extended away from fixed points of the RG, and used
to construct a function on the space of field theories
that monotonically decreases throughout RG flows and coincides
with the usual central charge $a$  at fixed points. This
extension has other uses as well, as we will demonstrate
with a few examples.

To illustrate the method, in the next section we discuss
the case of supersymmetric non-abelian gauge theory with
vanishing superpotential. In section 3 we include the effect
of superpotentials. Section 4 contains a discussion of some
generalizations of the results of sections 2 and 3 to situations
where some of the assumptions do not apply. 

\newsec{Supersymmetric gauge theories with vanishing superpotential}

Consider a supersymmetric gauge theory with gauge group $G$ and chiral
superfields $\Phi_i$ in the representations $r_i$ of the gauge group.
One can choose a basis of generators of the gauge group in the 
representation $r$, $T^a$, such that 
\eqn\quadcas{{\rm Tr}_r (T^a T^b)=T(r)\delta^{ab}~.}
The invariant \quadcas\ corresponding to the adjoint representation 
will be denoted by $T(G)$. For example, for $G=SU(N_c)$,
$T({\rm fundamental})=1/2$,  $T({\rm adjoint})=T(G)=N_c$.

The $\beta$-function for the gauge coupling $g$, or more conveniently for
$\alpha=g^2/4\pi$ is \refs{\NovikovUC,\ShifmanZI}
\eqn\betal{\beta(\alpha)=-{\alpha^2\over2\pi}{3T(G)-\sum_i T(r_i)(1-\gamma_i(\alpha))\over
1-{\alpha\over2\pi}T(G)}~.}
Here $\gamma_i$ is the anomalous dimension of $\Phi_i$; at fixed points of the RG, the
scaling dimension of $\Phi_i$ is given by
\eqn\dimgam{\Delta(\Phi_i)=1+{1\over2}\gamma_i~.}
At weak coupling one has 
\eqn\gamial{\gamma_i(\alpha)=-{\alpha\over\pi}C_2(r_i)+O(\alpha^2)}
where
\eqn\quadcas{C_2(r)={|G|\over |r|} T(r)~.}
$|G|$ and $|r|$ are the dimensions of the group $G$ and the representation
$r$, respectively. Eq. \betal\ implies that $\alpha$ is (marginally) relevant
at weak coupling if the theory is asymptotically free,
\eqn\asfrcon{3T(G)-\sum_i T(r_i)>0~.}
In this case $\alpha$ grows as the distance scale increases,
and the theory approaches a non-trivial  fixed point at long distances. 
At that fixed point, where $\alpha=\alpha^*$, one must have 
\eqn\fpcond{3T(G)-\sum_iT(r_i)(1-\gamma_i(\alpha^*))=0~,}
assuming, as we will do below, that $\alpha^*$ is sufficiently small 
so that the denominator of \betal\ never vanishes along the RG flow. 

The  condition \fpcond\ has a well known interpretation
in terms of the R-charges at the IR fixed point of the flow. Using
\dimgam\ and the relation between the scaling dimension $\Delta$
and superconformal $U(1)_R$ charge, $R$,
\eqn\delrrel{\Delta={3\over2} R}
one can rewrite \fpcond\ as
\eqn\anomcond{T(G)+\sum_i T(r_i)(R_i(\alpha^*)-1)=0~.}
This equation is also the condition that the R-symmetry
with $R(\Phi_i)=R_i(\alpha^*)$ be anomaly free and thus
conserved throughout the RG flow labeled by $\alpha$. 
We conclude that supersymmetric gauge theory with any matter
that satisfies the asymptotic freedom condition \asfrcon\ has the
property  that the IR $U(1)_R$ is conserved throughout the RG
flow, as long as the NSVZ $\beta$-function \betal\ is reliable.

In general, eq. \anomcond\ does not determine the R-charges
at the IR fixed point uniquely, but since the IR $U(1)_R$ is a
symmetry of the full theory, we can use the results of
\IntriligatorJJ\ to determine the $R_i$. The anomaly \aaa\ takes
in this case the form
\eqn\formanom{
a(R_i)=2|G|+\sum_i |r_i|\left[ 3(R_i-1)^3-(R_i-1)\right]}
where the first contribution is due to the gauginos and the second to the
quarks in the chiral multiplets.  The IR superconformal  $U(1)_R$ charges 
can be found by (locally) maximizing  $a$, \formanom, subject to the
constraint \anomcond. The value of the central charge $a_{IR}$  is obtained
by substituting the resulting R-charges into \formanom. To compute $a_{UV}$
one notes that at short distances the theory is free, and all the R-charges are given
by their free field value,  $R_i=2/3$.

It is natural to ask whether the prediction of the
$a$-theorem \athm\ is satisfied in this case. We will next show
that \athm\ indeed holds under the assumptions stated above.
In order to prove this, it is convenient to start with a generalization\foot{In
a slight abuse of notation, we will denote this generalization by $a$ as well.}
of \formanom\ that takes into account the constraint \anomcond:
\eqn\genera{ a(R_i,\lambda)=
2| G|+\sum_i |r_i|\left[ 3(R_i-1)^3-(R_i-1)\right]-
\lambda\left[T(G)+\sum_i T(r_i)(R_i-1)\right]~.}
One can think of the new variable $\lambda$ as a Lagrange
multiplier imposing the constraint  \anomcond. The procedure
for finding the IR $U(1)_R$ described in the previous paragraph 
is equivalent to the following. 

Find the local maximum of $a(R_i,\lambda)$ with respect to the $R_i$,
keeping $\lambda$ fixed and arbitrary. Note, in particular, that all the
$R_i$ are taken to be independent - we do not impose the constraint
\anomcond. At the local maximum, the $R_i$ are functions of $\lambda$.
Substituting their values into \genera, one finds $a=a(R_i(\lambda),\lambda)$.
In order to impose the constraint \anomcond\ one now extremizes $a$ with
respect to $\lambda$, by imposing
\eqn\extalambda{{da(R_i(\lambda),\lambda)\over d\lambda}=0~.}
This fixes $\lambda$ and thus $R_i(\lambda)$. We will denote the value of
$\lambda$ that solves \extalambda\ by $\lambda^*$. We will see below that
$R_i(\lambda^*)$ satisfy the constraint \anomcond; in general, the procedure
described here is equivalent to maximizing \formanom\ subject to the constraint
\anomcond.

So far we simply restated the original determination of the R-charges
at the IR fixed point of a super-Yang-Mills (SYM) theory in a slightly
different way. However, one might wonder whether there is some further
information in the $\lambda$ dependence of the generalized central charge
$a(R_i(\lambda),\lambda)$. An interesting fact is that for $\lambda=0$, the
generalized central charge \genera\ is simply that of free field theory without
any gauge interaction, and maximizing it with respect to the $R_i$ gives the
free field values $R_i=2/3$. Thus, as we vary $\lambda$ between $0$ and
$\lambda^*$, the central charge $a(R_i(\lambda),\lambda)$ varies continuously
between the UV, free field theory, value, and the IR value corresponding to the
interacting fixed point.

It is now easy to prove that \athm\ holds in this case. One has
\eqn\ddaallaa{{da(R_i(\lambda),\lambda)\over d\lambda}=
{\partial a\over\partial\lambda}+
\sum_i{\partial a\over\partial R_i}{\partial R_i\over\partial\lambda}~.}
The second term on the r.h.s. vanishes by construction, since $R_i(\lambda)$
is found by solving the equation $\partial_{R_i} a=0$. 
The first term can be read off  \genera:
\eqn\deralam{{da\over d\lambda}=- \left[T(G)+\sum_i T(r_i)(R_i(\lambda)-1)\right]~.}
At $\lambda=0$, $R_i(0)=2/3$ and the r.h.s. is nothing but the coefficient in 
the one loop $\beta$-function which is negative by assumption (see \asfrcon). 
It remains negative for all positive $\lambda$ up to $\lambda=\lambda^*$ 
(as we will see momentarily, $\lambda^*$ is positive), where it vanishes.
We see that $ a(R_i(\lambda),\lambda)$ is a decreasing function of $\lambda$ between
$\lambda=0$, which corresponds to the UV fixed point, and $\lambda=\lambda^*$,
which corresponds to the IR fixed point. Hence, $a_{UV}>a_{IR}$, in accordance with 
\athm. Note also that $\lambda^*$, which is by definition an extremum of 
$a(R_i(\lambda),\lambda)$ \extalambda, is in fact a (local) minimum of that function.
  
It is useful to write explicitly the solution of the equations described above.
Maximizing  \genera\ with respect to $R_i$ leads to 
\eqn\formri{R_i(\lambda)=1-{1\over3}
\left[1+{\lambda T(r_i)\over |r_i|}\right]^{1\over2}~.}
$\lambda^*$ is the solution of the constraint \anomcond,
\eqn\detlambda{T(G)+\sum_i T(r_i)(R_i(\lambda^*)-1)=0~.}
Using \formri\ it is easy to see that $\lambda^*$ is positive when the 
theory is asymptotically free \asfrcon. 

Although $\lambda$ was originally introduced in eq. \genera\ as a
Lagrange multiplier, its properties are reminiscent of those of the gauge
coupling $\alpha$.  We next argue that $\lambda$ and $\alpha$ are in fact 
related.

To see that, it is useful to recall the following well known story.
If the matter content of a gauge theory is such that the theory is just 
barely asymptotically free, then the IR fixed point can be studied 
perturbatively in $\alpha$ \refs{\GrossCS,\BanksNN}. For supersymmetric 
theories, this is the case when the coefficient in the one loop 
$\beta$-function, $3T(G)-\sum_i T(r_i)$ is positive \asfrcon, but very 
small. To find the perturbative fixed point, one then looks for solutions 
of the fixed point condition \anomcond, where the R-charges $R_i(\alpha)$
are related to the anomalous dimensions $\gamma_i(\alpha)$ via the
relation (see \dimgam, \delrrel) 
\eqn\relrigami{3R_i(\alpha)=2+\gamma_i(\alpha)~.} 
Both the anomalous dimensions, $\gamma_i(\alpha)$, and the R-charges,
$R_i(\alpha)$, have an expansion in $\alpha$. For the R-charges one has
\eqn\rialpha{R_i(\alpha)={2\over3}+R_i^{(1)}\alpha+R_i^{(2)}\alpha^2+
\cdots}
where the $R_i^{(n)}$ can be computed (in a particular scheme, or parametrization
of coupling space) by performing loop calculations in the gauge theory.
The value of $\alpha$ at the IR fixed point is obtained by plugging \rialpha\
into \anomcond\ and solving the resulting equation for $\alpha^*$. If
$\epsilon\equiv 3T(G)-\sum_i T(r_i)$ is very small and positive, one is led
to an expansion of $\alpha^*$ (and thus of $R_i(\alpha^*)$ \rialpha), in a
power series in $\epsilon$.

The results of IW reviewed above allow one to compute the IR R-charges 
$R_i(\alpha^*)$ directly, by using $a$-maximization. In the
description of the method of IW given above, the $R_i$ are functions of
$\lambda$ \formri. They have a Taylor expansion analogous to \rialpha,
and the value of $\lambda$ at the IR fixed point is obtained by solving
\detlambda, which has the same form as the fixed point condition \anomcond.

The similarity of the way $\alpha$ and $\lambda$ enter the problem leads
us to propose that they should be identified, in the sense that both 
parametrize the space of gauge couplings. Thus, they should be related
as follows\foot{One expects that $\lambda$ should be a monotonic function
of $\alpha$, at least for $0\le \alpha\le \alpha^*$.}:
\eqn\lamcal{\lambda=A_1\alpha+A_2\alpha^2+A_3\alpha^3+\cdots}
where $A_1, A_2,A_3\cdots$ are numerical coefficients to be determined.
Analytic redefinitions of the form
$\alpha=\tilde\alpha+a\tilde\alpha^2+b\tilde\alpha^3+\cdots$, which relate 
different renormalization schemes, change the coefficients $A_2,A_3,\cdots$
but the leading coefficient $A_1$ is invariant under such redefinitions.
It can be determined by studying the leading weak coupling behavior of
the theory. Using \gamial, one finds
\eqn\rigauge{R_i(\alpha)={2\over3}-{\alpha\over3\pi}{|G|\over|r_i|}T(r_i)+O(\alpha^2)~.}
On the other hand, expanding \formri\ one has
\eqn\rriill{R_i(\lambda)={2\over3}-{\lambda\over6}{T(r_i)\over |r_i|}+O(\lambda^2)~.}
Comparing the two, one concludes that
\eqn\rellamal{\lambda={2\alpha\over\pi}|G|+O(\alpha^2)~.}
Thus, the relation between $\alpha$ and $\lambda$ is independent 
of the particular representation $r_i$. This had to be the case, 
given the fact that we are solving essentially the same equations \anomcond, 
\detlambda, and expanding in a power series in $\alpha$ and $\lambda$, 
respectively. Nevertheless, the agreement supports the relation between
$\alpha$ and $\lambda$ proposed above. 

We see that the $a$-maximization procedure of \IntriligatorJJ\ together 
with the postulate that $\lambda$ is proportional to $\alpha$ at weak 
coupling \lamcal\ predict the dependence of the anomalous dimensions 
on the representation $r_i$ given in eq. \gamial\ (\ie\ the fact that 
$\gamma_i(\alpha)\propto \alpha T(r_i)/|r_i|$ to leading order in $\alpha$). 
This is a non-trivial result which is normally derived by evaluating one 
loop diagrams in the SYM theory. 

The above discussion can be extended to higher loops (\ie\ higher orders
in $\alpha$, $\lambda$). One can expand $R_i$ \formri\ in a power series
in $\lambda$, and use \relrigami\ and \lamcal\ to find constraints on the
anomalous dimensions $\gamma_i(\alpha)$ at higher loop order. For example,
to order $\alpha^2$ one finds 
\eqn\twoloop{\gamma_i(\alpha)=-{\alpha\over\pi}C_2(r_i)+
{\alpha^2\over2\pi^2}\left(C_2(r_i)\right)^2-
{A_2\over2|G|}\alpha^2C_2(r_i)+O(\alpha^3)~.}
The two loop $(O(\alpha^2))$ term in the anomalous dimension is 
scheme-dependent. We see that the term that goes like $(C_2(r_i))^2$ 
is in fact scheme-independent, and the scheme dependence appears in 
the term proportional to $C_2(r_i)$ via the coefficient $A_2$ \lamcal. 
The two loop anomalous dimensions in $N=1$ SYM theories can be found
for example in \JackCN, which also contains references to the
original literature on the subject. One can readily check that
the structure of the two loop anomalous dimension is as in \twoloop,
and the coefficient of $(C_2(r_i))^2$ agrees. The coefficient
of $C_2(r_i)$ in \JackCN\ can be used to compute $A_2$ for the 
particular renormalization scheme used there. 

Some comments are appropriate at this point:
\item{(1)} The result \formri\ shows that the R-charges, and thus 
the scaling dimensions \delrrel, monotonically decrease along the RG
flow. This is not a general feature of four dimensional QFT. We will 
see in the next section that superpotential terms often have the 
opposite effect. 
\item{(2)} In analogy to two dimensional QFT, one might want to
require that the central charge $a(R_i(\lambda),\lambda)$ satisfy 
the relation 
\eqn\gradflow{{da\over d\alpha}=\beta(\alpha)G(\alpha)}
where $G(\alpha)$ is the metric on coupling space. 
Using \betal, \deralam\ and \lamcal\ one finds that
this metric is
\eqn\metcop{G(\alpha)={d\lambda\over d\alpha}
{2\pi\over 3\alpha^2}\left[1-{\alpha\over2\pi}T(G)\right]
\simeq {4|G|\over 3\alpha^2}}
where in the last approximate equality we used \rellamal\
to evaluate $G$ for small $\alpha$. In this metric, the distance
to the free field fixed point $\alpha=0$ from any finite $\alpha$
is infinite -- it goes like $\int_0 d\alpha\sqrt{G(\alpha)}
\simeq \int_0 {d\alpha\over\alpha}$. 
\item{(3)} The relation between $\lambda$ and $\alpha$ in the
vicinity of the UV fixed point, \rellamal, shows that $\lambda$
must be positive. Similar restrictions on Lagrange multipliers
associated with superpotentials will play a role in our discussion
later.
\item{(4)} The authors of \AnselmiYS\ proposed a different
interpolating central charge, which in our notation is given by
\eqn\altaa{\tilde a(\lambda)=2|G|+\sum_i|r_i|
\left[3(R_i(\lambda)-1)^3-(R_i(\lambda)-1)\right]~.}
This charge decreases as well under RG flow:
\eqn\altbb{{d\tilde a\over d\lambda}=
\lambda\sum_i T(r_i){dR_i\over d\lambda}=
-{\lambda\over6}\sum_i{T^2(r_i)\over |r_i|}
\left(1+{\lambda T(r_i)\over |r_i|}\right)^{-{1\over2}}~,}
where we have used \formri.

We end this section with two examples. The first is $N=1$ supersymmetric
QCD (SQCD). The gauge group is $G=SU(N_c)$; the matter consists of $N_f$ 
chiral superfields $Q_i$, $\widetilde Q^i$ ($i=1,\cdots, N_f$) transforming 
in the ${\bf N_c}$, ${\bf \bar N_c}$ of $SU(N_c)$, respectively. 

This theory is asymptotically free for $N_f<3N_c$. The results
of \IntriligatorJJ\ are not needed to determine the superconformal
$U(1)_R$ of the IR theory in this case. The R-charges $R_Q$, 
$R_{\widetilde Q}$ must be taken to be equal, due to the $Z_2$ symmetry 
of exchanging $Q$ and $\widetilde Q$. The anomaly constraint \anomcond\ 
takes the form $N_c+N_f(R_Q-1)=0$, with the solution 
\eqn\rqncnf{R_Q=R_{\widetilde Q}=1-{N_c\over N_f}~.}
It is known \SeibergPQ\ that this result is correct for 
$3N_c/2\le N_f\le 3N_c$. For $N_f<3N_c/2$, the NSVZ $\beta$-function
breaks down in a way that is difficult to understand from the
present perspective. The physics is then believed to be described
by the Seiberg-dual theory, which has gauge group $G_m=SU(N_f-N_c)$.
We will have nothing new to say about this here; thus, we will restrict
our attention to the region $N_f\ge 3N_c/2$. 

The UV and IR central charges $a_{UV}$ and $a_{IR}$ 
are:
\eqn\auvir{\eqalign{
a_{UV}=&2(N_c^2-1)+2N_fN_c{2\over9}~,\cr
a_{IR}=&2(N_c^2-1)+2N_fN_c\left[3(R_Q-1)^3-(R_Q-1)\right]~,\cr
}}
with $R_Q$ given by \rqncnf. 
It can be checked directly that $a_{UV}>a_{IR}$ in the region under
discussion \refs{\AnselmiAM,\AnselmiYS} but one can also apply
our reasoning above. Eq. \formri\ takes in this case the form
\eqn\rqlam{R_Q(\lambda)=1-{1\over3}
\left(1+{\lambda\over2 N_c}\right)^{1\over2}~.}
Substituting this into \genera\ one finds 
\eqn\aqlam{a(R_Q(\lambda),\lambda)=2(N_c^2-1)+
{4\over9}N_fN_c\left(1+{\lambda\over2 N_c}\right)^{3\over2}
-\lambda N_c~.}
At $\lambda=0$ this agrees with the UV value \auvir.
Differentiating with respect to $\lambda$ one finds, as in
\deralam:
\eqn\dalam{{da\over d\lambda}=-N_c+{N_f\over3}
\left(1+{\lambda\over2 N_c}\right)^{1\over2}
=-N_c+N_f(1-R_Q(\lambda))~.}
We see that $a(R_Q(\lambda),\lambda)$ decreases as $\lambda$
grows. $\lambda^*$, the solution of \detlambda, is given here
by 
\eqn\lambstar{{\lambda^*\over 2N_c}=\left({3N_c\over N_f}\right)^2
-1}
which is indeed positive for $N_f<3N_c$. 

A point that is nicely illustrated by the SQCD example is that 
$a$-maximization is useful beyond the determination of the IR
$U(1)_R$ charge. As discussed above, the fact that the IR
R-charges are given by \rqncnf\ does not require it.
Instead, $a$-maximization is used here to continuously
interpolate between the UV and IR fixed points and define 
a monotonically decreasing function throughout the RG flow,
\aqlam.

As mentioned above, the discussion must break down for $N_f<3N_c/2$.
Presumably, this is due to the fact that in this range, during
the RG flow $\alpha$ reaches the value at which the denominator
in \betal\ has a pole, after which the NSVZ $\beta$-function
\betal\ might become unreliable. If we knew the relation 
between $\lambda$ and $\alpha$ \lamcal, we would be able 
to check this quantitatively, but unfortunately the precise
relation between the two is not known at present. 

Our second example is adjoint SQCD, a model with gauge group
$G=SU(N_c)$, $N_f$ flavors of fundamentals $Q_i$, $\widetilde Q^i$
as before, but now with an extra chiral superfield $X$ transforming
in the adjoint representation of the gauge group. This model is
asymptotically free for $N_f<2N_c$, and is believed to flow in the
IR to a non-trivial fixed point for all $N_f$ in this range.

As before, $Q_i$ and $\widetilde Q^i$ have the same R-charge $R_Q$,
while $X$ has R-charge $R_X$. The anomaly constraint \anomcond\ is
\eqn\anrqrx{N_fR_Q+N_cR_X=N_f~.}
It does not fix $R_Q$, $R_X$ uniquely and one has to use the results
of \refs{\IntriligatorJJ, \KutasovIY} to do that. 

Our general discussion gives rise in this case to the following
$\lambda$-dependent R-charges:
\eqn\rqxlam{\eqalign{
R_Q(\lambda)=&1-{1\over3}\left(1+{\lambda\over 2 N_c}\right)^{1\over2}\cr
R_X(\lambda)=&1-{1\over3}\left(1+{\lambda N_c\over N_c^2-1}\right)^{1\over2}~.\cr
}}
The IR fixed point is at $\lambda=\lambda^*>0$, for which the R-charges
\rqxlam\ satisfy \anrqrx. One can check that the resulting IR R-charges
are consistent with those given in \refs{\IntriligatorJJ,\KutasovIY}.

Adjoint SQCD is an example where our construction interpolates
smoothly between two fixed points in each of which one needs to maximize
$a$ with respect to the $R_i$. As in the general case, \deralam, the central
charge monotonically decreases throughout the flow. 

One other comment that needs to be made here is that, as pointed out in
\KutasovIY, when the R-charges of $Q$, $\widetilde Q$ or $X$ drop below 
$1/3$, one needs to modify the procedure of IW to take into account 
unitarity constraints. We will not attempt to incorporate these 
corrections here; it would be interesting to do so.

\newsec{Models with non-vanishing superpotentials}

So far we only discussed supersymmetric non-abelian gauge interactions.
In this section we will include the effects of superpotentials. 

\subsec{Chiral superfields with non-zero superpotentials}

In this subsection we consider theories with no gauge fields.
We expect no interesting dynamics in this case, since such
theories are always IR free in four dimensions (at least at
weak coupling), but we can still apply our methods to them
and see what we get. This discussion will also serve as preparation
for the case with gauge interactions.

Let $\Phi$ be a chiral superfield with standard kinetic term,
and consider the effect of various polynomial superpotentials.
For example, take
\eqn\wmphi{W=m\Phi^2~.}
This is a mass term, and we expect it to lead to an RG flow
with a trivial IR fixed point. From the point of view of our
analysis one can include the effect of \wmphi\ as follows.

Let $R_\phi$ be the R-charge of $\Phi$. Consider the generalized 
central charge
\eqn\arphil{a(R_\phi,\lambda)=3(R_\phi-1)^3-(R_\phi-1)+
\lambda(R_\phi-1)~.}
At $\lambda=0$, this is just the central charge
of a free massless chiral superfield; the local
maximum of $a(R_\phi,0)$ is at $R_\phi=2/3$, the
free value \IntriligatorJJ. $\lambda$ is a Lagrange
multiplier enforcing the condition $R_\phi=1$, which
is necessary for the R-symmetry to be preserved by the
superpotential \wmphi. We can again find the local maximum
of \arphil\ at fixed $\lambda$. This occurs at
\eqn\rphimax{R_\phi=1-{1\over3}\sqrt{1-\lambda}~.}
Substituting back into \arphil\ we find
\eqn\aarrll{a(R_\phi(\lambda),\lambda)={2\over9}
(1-\lambda)^{3\over2}~.}
We see that $a$ interpolates between the value $a=2/9$
(corresponding to a free superfield) at $\lambda=0$,
and $a=0$ at $\lambda=1$ where ${da\over d\lambda}=0$
(see \extalambda). Between the two values, $a$ monotonically
decreases, since
\eqn\mondec{{da(R_\phi(\lambda),\lambda)\over d\lambda}
=R_\phi(\lambda)-1}
which is negative for all $\lambda<1$
(compare to \ddaallaa, \deralam). One can again argue
that $\lambda$ in \arphil, which starts its life as a
Lagrange multiplier, actually parametrizes the coupling
space \wmphi. In the field theory this space is labeled
by $m/E$, where $E$ is the energy scale at which we are
working. The ``coupling'' $m/E$ varies between $0$ and
$\infty$. In the description \aarrll\ this is translated
into the dependence on $\lambda$, which varies between 
$0$ and $1$. 

Note also that the discussion above provides
a simple example in which the R-charge increases under
RG flow. This should be contrasted with gauge interactions, 
which lower the R-charges, see \formri. 

Next we discuss the case of a cubic superpotential
\eqn\cubsup{W=\mu\Phi^3~.}
This interaction is marginally irrelevant, so one does
not expect to find any interacting fixed points at 
$\mu\not=0$. In this case, the central charge takes the form
\eqn\acubic{a(R_\phi,\lambda)=3(R_\phi-1)^3-(R_\phi-1)+
\lambda(R_\phi-{2\over3})~.}
The local maximum in $R_\phi$ is again at \rphimax, and
\eqn\amaxcub{a(R_\phi(\lambda),\lambda)={2\over9}
(1-\lambda)^{3\over2}+{1\over3}\lambda~.}
Imposing \extalambda\ we find 
\eqn\maxalam{{da(R_\phi(\lambda),\lambda)\over d\lambda}
=-{1\over3}\sqrt{1-\lambda}+{1\over3}=0~.}
The only solution is $\lambda=0$, the original trivial
fixed point with $\mu=0$ \cubsup. Note in particular that
$a$ \amaxcub\ ``knows'' about the fact that the cubic
superpotential \cubsup\ is marginally irrelevant. In 
\amaxcub\ this is the statement that while the term linear
in $\lambda$ vanishes, the quadratic term is positive, so
$\lambda=0$ is a local minimum of $a$. 

More generally, one can consider a superpotential of the
form $W=\mu\Phi^n$. Repeating the analysis for this case 
one finds that $R_\phi(\lambda)$ is given by \rphimax\ 
and the central charge is
\eqn\arphinone{a(R_\phi(\lambda),\lambda)={2\over9}
(1-\lambda)^{3\over2}+\lambda{n-2\over n}~.}
The stationarity condition is
\eqn\daleqz{{da\over d\lambda}
=-{1\over3}\sqrt{1-\lambda}+{n-2\over n}=0~.}
For $n>3$, this equation has a solution, $\lambda^*$, but it
is negative. We have seen before that the Lagrange
multiplier $\lambda$ must have a particular sign, both in
the gauge theory example of section 2, where this sign is
correlated with that of $\alpha$ (see \rellamal), and in the case
of the mass deformation described earlier in this subsection. The
general rule is that the sign of $\lambda$ is constrained to be
such that when we perturb a fixed point by a {\it relevant}
operator, $a(\lambda)$ should {\it decrease} when we turn on 
$\lambda$, and vice-versa. 
In other words, if we choose $\lambda$ to be positive by convention,
it must be that $\partial_\lambda a<0$ at $\lambda=0$ for a 
relevant perturbation, and $\partial_\lambda a>0$ for an irrelevant
one. This is why in section 2, $\lambda$ was constrained to be
positive (see \deralam) and it also implies that in \daleqz\
$\lambda$ should be taken to be positive. 

Thus, the solutions of \daleqz\ with $\lambda<0$ should be discarded,
and we do not find any fixed points with non-zero $\lambda$ (or $\mu$)
when we turn on an irrelevant superpotential, as one would expect. 

The only remaining case is a linear superpotential, $W=\mu\Phi$. 
The stationarity condition is in this case
\eqn\linstat{-{1\over3}\sqrt{1-\lambda}-1=0~.}
It has no solutions, in agreement with the
fact that for a linear superpotential, the condition
for a supersymmetric ground state, $W'=0$ has no solutions.

We can also study models in which the superpotential
is a more general polynomial
\eqn\wgenerk{W=\sum_{n=2}^k a_n\Phi^n~.}
The most important new effect here is that the R-symmetry
is in general completely broken throughout the RG flow.
For example, if $a_2\not=0$ and we focus on the vacuum
with $\langle\Phi\rangle=0$, the R-charge of $\Phi$ still 
approaches $1$ in the IR, as in \rphimax\ in the limit 
$\lambda\to 1$, but now this symmetry is present only 
at the IR fixed point, where one can neglect the higher 
powers of $\Phi$ in \wgenerk.

One can generalize our analysis of the different cases above and
write the central charge corresponding to \wgenerk\ as
\eqn\genalamn{a(R_\phi,\lambda_2,\cdots,\lambda_k)=
3(R_\phi-1)^3-(R_\phi-1)+\sum_{n=2}^k\lambda_n(R_\phi-{2\over n})~,}
where $(\lambda_2,\cdots,\lambda_k)$ parametrize the space of
couplings $(a_2,\cdots, a_k)$ in \wgenerk. After solving for 
$R_\phi$ as a function of $(\lambda_2,\cdots, \lambda_k)$ one 
gets a central charge defined on the space of $\lambda$'s, 
$a(R_\phi(\lambda_2,\cdots,\lambda_k),\lambda_2,\cdots,\lambda_k)$.
We will not pursue this here since most of the couplings $\lambda_n$
are irrelevant. Instead we will move on to the case where both gauge
interactions and superpotentials are present.

\subsec{Models with gauge interactions}

In the last subsection we have seen that in the absence 
of non-abelian gauge fields there is little interesting 
dynamics, essentially because the only relevant perturbation
of a massless free chiral superfield is a mass term.

The situation is different in non-abelian gauge theories. 
As we have seen in section 2, gauge interactions decrease 
the scaling dimensions and R-charges of chiral superfields 
(see \formri); thus, it often happens that the IR fixed 
point of a gauge theory has relevant perturbations that 
did not exist at the (free) UV fixed point. In this subsection 
we will study the effect of these perturbations on the
discussion of section 2.

The basic setup will be the same as in section 2. We have an 
$N=1$ SYM theory with gauge group $G$ and chiral superfields 
$\Phi_i$ in representations $r_i$. We assume that the theory 
is asymptotically free and flows in the IR to a fixed point
which is described by the NSVZ $\beta$-function \betal\ 
(\ie\ satisfies \fpcond, \anomcond). We further assume that 
at the IR fixed point of this gauge theory, the gauge invariant
chiral operator 
\eqn\chirop{\OO=\prod_i \Phi_i^{n_i}}
is relevant, \ie\
\eqn\relcond{\sum_i n_iR_i(\alpha^*)<2~.}
We thus can add to the Lagrangian the superpotential
\eqn\addsup{W=\mu\OO}
which drives the theory to a new infrared fixed point, which we will refer to
as $IR'$. At that new fixed point, the R-charges $R_i$ must satisfy the constraint
\eqn\irrcond{\sum_i n_iR_i=2~.}
We will furthermore assume that the anomaly condition \anomcond\ is compatible
with \irrcond, so that the R-symmetry at the new fixed point $IR'$, which satisfies both
\anomcond\ and \irrcond, is preserved throughout the RG flow $UV \to IR \to IR'$.

We would like to show that under these conditions
\eqn\newineq{a_{IR}>a_{IR'}~.}
To do that, we consider the generalized central charge 
\eqn\genasup{\eqalign{ &a(R_i,\lambda,\lambda')=
2| G|+\sum_i |r_i|\left[ 3(R_i-1)^3-(R_i-1)\right]\cr
-&\lambda\left[T(G)+\sum_i T(r_i)(R_i-1)\right]
-\lambda'\left(2-\sum_i n_iR_i\right)~.\cr}}
$\lambda$, $\lambda'$ are again Lagrange multipliers
enforcing the constraints \detlambda\ and \irrcond.
As discussed in the last subsection, they are both
non-negative. 

Following the same logic as in our previous discussions,
we find the local maximum of \genasup\ with respect to
the $R_i$, at fixed $(\lambda,\lambda')$. This occurs at
\eqn\rillprime{R_i(\lambda,\lambda')=
1-{1\over3}\left(1+{\lambda T(r_i)-
\lambda' n_i\over |r_i|}\right)^{1\over2}~.}
Substituting this back into \genasup\ we arrive
at a central charge $a(R_i(\lambda,\lambda'),\lambda,\lambda')$,
(which we will denote by $a(\lambda,\lambda')$ for brevity)
defined on the space of theories labeled by the gauge coupling
(which is related to $\lambda$, as discussed in section 2), and the
superpotential coupling $\mu$ \addsup, which is related to $\lambda'$.

The derivation of eq. \deralam\ can be repeated for this case. One finds:
\eqn\ineqprime{\eqalign{
{\partial a\over \partial\lambda}=&- \left[T(G)+\sum_i T(r_i)(R_i(\lambda,\lambda')-1)\right]~,\cr
{\partial a\over \partial\lambda'}=&- \left[2-\sum_i n_iR_i(\lambda,\lambda')\right]~.\cr
}}
The three fixed points discussed above are described as follows in terms of $a(\lambda,\lambda')$.
The free UV fixed point corresponds to $\lambda=\lambda'=0$. The IR fixed point of the RG flow
with vanishing superpotential is at $\lambda'=0$, $\lambda=\lambda_1^*$, with $\lambda_1^*$
determined by solving
\eqn\minlam{{\partial a(\lambda,0)\over\partial\lambda}=0~.}
The fixed point $IR'$ to which the system is driven by \addsup\ is at
$\lambda=\lambda_2^*$, $\lambda'=\lambda_2'^*$ which are obtained by minimizing
$a(\lambda,\lambda')$ with respect to both $\lambda$ and $\lambda'$:
\eqn\secmin{{\partial a\over\partial\lambda}=
{\partial a\over\partial\lambda'}=0\;\;\;{\rm at}
\;\;\;(\lambda,\lambda')=(\lambda_2^*,\lambda_2'^*)~.}
In order to derive \newineq, it is useful to recall 
some aspects of the analysis of section 2. Before 
turning on the superpotential \addsup\ (\ie\ at 
$\lambda'=0$), we saw that $a(\lambda,0)$ has a local
minimum at $\lambda=\lambda_1^*$. Turning on $\lambda'$ 
should not change the picture qualitatively. As we see 
from the first line of \ineqprime, 
$\partial_\lambda a(\lambda,\lambda')$ is negative for 
$0\le \lambda <\lambda^*(\lambda')$, where 
$\lambda^*(\lambda')$ is obtained by solving the equation
\eqn\solveq{{\partial a(\lambda,\lambda')\over\partial\lambda}=0~.}
For $\lambda'=0$ we have $\lambda^*(0)=\lambda_1^*$. 
Now consider following a trajectory from the fixed 
point $IR$ to $IR'$, \ie\ from $(\lambda,\lambda')=(\lambda_1^*,0)$ to
$(\lambda,\lambda')=(\lambda_2^*,\lambda_2'^*)$, along the curve $\lambda^*(\lambda')$.
Along this curve one has
\eqn\derag{{da(\lambda^*(\lambda'),\lambda')\over d\lambda'}
={\partial a\over\partial\lambda'}
+{\partial a\over\partial\lambda}{\partial \lambda^*\over\partial\lambda'}~.}
As in \ddaallaa, the second term vanishes since along the trajectory in question
$\partial_\lambda a=0$ \solveq. The first term is given by \ineqprime. Thus, we conclude
that
\eqn\derapp{{d a(\lambda^*(\lambda'),\lambda')\over d\lambda'}=
- \left[2-\sum_i n_iR_i(\lambda^*(\lambda'),\lambda')\right]~.}
The r.h.s. is negative at $\lambda'=0$ \relcond\ and it remains negative
all the way to $\lambda'=\lambda_2'^*$ where it vanishes.

Eq. \derapp\ implies the inequality \newineq. Along the trajectory
$\lambda^*(\lambda')$ connecting the fixed points $IR$ and $IR'$, 
the central charge $a$ is monotonically decreasing, so $a_{IR}>
a_{IR'}$. This concludes the proof of the inequality \newineq.

Note that we assumed above that $\lambda_2'^*$ is positive.
The reason for that is that, as explained in the previous subsection,
$\lambda'$ is by definition non-negative. Thus, any fixed point obtained
by adding the relevant perturbation \addsup\ must give rise to a critical
point of $a(\lambda,\lambda')$ \secmin\ at positive $\lambda$ and $\lambda'$.
If there are no solutions of eq. \secmin\ with $\lambda_2'^*>0$, then we
conclude that this perturbation breaks supersymmetry, like the linear
superpotential discussed in the previous subsection. 

There are many additional
aspects of the construction that are worth exploring. Since our primary 
interest was in establishing the hierarchy $a_{UV}>a_{IR}>a_{IR'}$, we
focused on a very specific trajectory in the space of couplings. Starting
at the UV fixed point, we turned on only the gauge coupling, went to the
IR fixed point, and only then turned on the relevant superpotential \addsup,
which drove the system to the fixed point $IR'$. 

One could study more general RG trajectories in coupling space, that
lead directly from $UV\to IR'$, without passing through $IR$. 
Such trajectories involve turning on both the gauge coupling
and $\mu$ \addsup, or both $\lambda$ and $\lambda'$, as one leaves
the UV fixed point. 

As mentioned in the beginning of this subsection, the operator
\chirop\ is in general irrelevant near the UV fixed point, and
only becomes relevant at some point along the flow of the gauge
coupling. Thus, if we turn on both the gauge coupling $\alpha$,
and $\mu$ \addsup, at first $\alpha$ will grow and $\mu$ decrease
along the RG flow, but at some point, when $\alpha$ passes a critical
value such that \chirop\ becomes relevant, $\mu$ will start growing
as well. 

The generalized central charge $a(\lambda,\lambda')$ provides a
good quantitative tool for studying such flows. The second line
of \ineqprime\ shows that for small $\lambda$, $\partial_{\lambda'}
a$ is positive, while beyond a critical value $\lambda_{cr}(\lambda')$
it becomes negative. Qualitatively, RG flows correspond to rolling in a
landscape with height function $a(\lambda,\lambda')$. It would be
nice to understand quantitatively the shape of the resulting
trajectories in the $(\lambda,\lambda')$ plane. 

An interesting logical possibility related to these trajectories is 
that one might think that it is possible that even if the operator 
$\CO$ \chirop\ is irrelevant at the IR fixed point of the gauge 
theory, the fixed point $IR'$ could still exist, and one could reach 
it without going through the $W=0$ IR fixed point. We will see later
that this possibility is not realized in an example that we study in
detail. A more general analysis of this issue will be left to future work.

Another natural generalization is to multi-step cascades, where
one flows from $UV\to IR^{(1)}\to IR^{(2)}\to IR^{(3)}\to\cdots$.
To go from $IR^{(n)}\to IR^{(n+1)}$ one turns on a superpotential
$W=\mu_n\CO_n$ that is relevant at the fixed point $IR^{(n)}$.
As long as the assumptions of our analysis are valid, \ie\ if the
R-symmetry under which $R(\CO_n)=2$ is a symmetry of the full RG
flow, one can repeat the analysis above, and it leads to the hierarchy
\eqn\hieraa{a_{UV}>a_{IR}^{(1)}>a_{IR}^{(2)}>a_{IR}^{(3)}>\cdots~.}
This can be shown by studying the generalized central charge 
\eqn\manyon{\eqalign{ &a(R_i,\lambda,\lambda_n)=
2| G|+\sum_i |r_i|\left[ 3(R_i-1)^3-(R_i-1)\right]\cr
-&\lambda\left[T(G)+\sum_i T(r_i)(R_i-1)\right]
-\sum_n\lambda_n\left[2-R(\CO_n)\right]~.\cr}}
Maximizing in $R_i$ leads to a central charge $a(\lambda,\lambda_n)$
which can be used as in equations \ineqprime\ -- \derapp\ to prove
\hieraa. 

Another application of our formalism is to the study of moduli spaces
of CFT's. A generic way to obtain a non-trivial moduli space of fixed 
points is to study a gauge theory of the sort discussed above, with a
superpotential
\eqn\supgener{W=\sum_n\mu_n\CO_n}
in a situation where the conditions
$T(G)+\sum_i T(r_i)(R_i-1)=0$ and $R(\CO_n)=2$ are not
linearly independent (viewed as functions of the $R_i$). Thus,
there exist constants $A$, $A_n$, not all of which are
zero, such that
\eqn\linrel{A\left[T(G)+\sum_i T(r_i)(R_i-1)\right]+
\sum_nA_n\left[2-R(\CO_n)\right]=0~.}
In this case, one generically expects (see \eg\ \LeighEP)
the IR fixed point of the flow to be a manifold labeled 
by the couplings of exactly marginal operators. The number 
of moduli is the same as the number of independent relations 
\linrel. 

Relations of the form \linrel\ have the following effect
on our discussion. One can use each such relation to express
one of the terms in square brackets on the second line of 
\manyon\ in terms of the others, and thus eliminate it from
the equation. This decreases the number of independent terms,
but the coefficients of the remaining ones still depend on
the original couplings $(\lambda$, $\lambda_n)$.
Thus, $a(\lambda,\lambda_n)$ is in this case independent
of some linear combinations of $\lambda$ and the $\{\lambda_n\}$.
These combinations are the moduli of the IR CFT. This establishes
in general that both the central charge $a$, and the R-charges
$R_i$ are independent of the moduli. 

Although we have presented the arguments of this section in terms of 
deformations of a free field fixed point by a combination of gauge 
interactions and superpotentials, it should be clear that one can in 
fact phrase them abstractly as follows. Let $A$ be an $N=1$ superconformal 
field theory, and $\OO$ a chiral operator, such that the superpotential 
\addsup\ corresponds to a relevant perturbation of the SCFT $A$, leading 
at long distances to SCFT $B$. Assume further that the $U(1)_R$ symmetry 
of the IR SCFT $B$ is a subgroup of the symmetry group of the full theory. 
Then $a_A>a_B$. 

To prove this, start with the central charge $a$ corresponding 
to a generic $U(1)_R$ symmetry at the fixed point $A$. Under the 
assumptions stated above, this central charge depends on some 
continuous parameters $(s_1,\cdots, s_n)$, which parametrize the 
particular $U(1)_R$ symmetry whose anomaly is being computed. 
The special $U(1)_R$ which belongs to the superconformal multiplet 
is obtained by locally maximizing $a(s_1,\cdots, s_n)$ with respect 
to the $s_i$ \IntriligatorJJ. The effect of the perturbation \addsup\ 
is captured by studying the generalized central charge
\eqn\agenerr{a(\lambda,s_1,\cdots, s_n)=a(s_1,\cdots, s_n)-\lambda
\left[2-R_\OO(s_1,\cdots, s_n)\right]~,}
where $R_\OO$ is the R-charge of the chiral operator $\OO$ under the symmetry
labeled by $(s_1,\cdots, s_n)$. Locally maximizing $a(\lambda,s_1,\cdots, s_n)$
with respect to $(s_1,\cdots, s_n)$ and solving for the $s_i$ as a function
of $\lambda$ leads to an effective central charge $a(\lambda)$,
which interpolates between the fixed point $A$ at $\lambda=0$, and the fixed point
$B$ at $\lambda=\lambda^*$, corresponding to a local minimum of $a(\lambda)$
at a positive value of $\lambda$. Between the two fixed points, $a$ is monotonically
decreasing, since
\eqn\adecreases{{da\over d\lambda}=-\left[2-R_\OO(s_1,\cdots, s_n)\right]~.}

To conclude this section we briefly discuss an example of a model
with a non-vanishing superpotential, a deformation of adjoint SQCD,
which was introduced at the end of section 2.

It is not difficult to check, using equations \anrqrx, \rqxlam,
that the R-charge of $X$ at the IR fixed point of adjoint SQCD
with vanishing superpotential goes like $N_f/N_c$ for small 
$N_f/N_c$, and in particular it can become arbitrarily small,
at least when $N_f$, $N_c$ are large (see \refs{\IntriligatorJJ,
\KutasovIY} for a more detailed discussion). Thus, if $N_f/N_c$ 
is small enough, the operator ${\rm tr} X^{k+1}$ is relevant at the IR 
fixed point, and we can add to the Lagrangian the superpotential 
\eqn\xkpone{W=g_k {\rm tr} X^{k+1}~,}
which is of interest in the study of Seiberg duality 
\refs{\KutasovVE-\KutasovSS}. The generalized central charge
\genasup\ is now:
\eqn\genarxq{\eqalign{
&a(R_X,R_Q,\lambda,\lambda')=2(N_c^2-1)+\cr
&(N_c^2-1)\left[3(R_X-1)^3-(R_x-1)\right]+
2N_fN_c\left[3(R_Q-1)^3-(R_Q-1)\right]-\cr
&\lambda\left[N_cR_X+N_f(R_Q-1)\right]-
\lambda'\left[2-(k+1)R_X\right]\cr
}}
Maximizing with respect to $R_X$ and $R_Q$ one finds
\rillprime
\eqn\rchxqq{\eqalign{
&R_X(\lambda,\lambda')=1-{1\over3}\left(1+{\lambda N_c-\lambda'(k+1)
\over N_c^2-1}\right)^{1\over2}\cr
&R_Q(\lambda,\lambda')=1-{1\over3}\left(1+{\lambda\over 2N_c}
\right)^{1\over2}\cr
}}
The superpotential \xkpone\ is relevant when 
$R_X(\lambda,\lambda')<2/(k+1)$. The non-trivial
fixed point $IR'$ is obtained by setting (see
\anrqrx, \ineqprime, \secmin, \rchxqq) 
\eqn\valrxq{\eqalign{
R_X(\lambda_2^*,\lambda_2'^*)=&{2\over k+1}\cr
R_Q(\lambda_2^*,\lambda_2'^*)=&1-{2N_c\over (k+1)N_f}~.\cr
}}
For example, for the case of a cubic superpotential, 
$k=2$, one finds
\eqn\cubiclam{\eqalign{
{\lambda_2^*\over 2N_c}=&\left({2N_c\over N_f}\right)^2-1~,\cr
\lambda_2'^*=&{\lambda_2^* N_c\over3}~.\cr
}}
Note that both $\lambda_2^*$ and $\lambda_2'^*$ are positive,
as implied by our general discussion, when the theory
is asymptoticall free, $N_f<2N_c$. 

For $k>2$, one finds
\eqn\lamgenk{\eqalign{
{\lambda_2^*\over 2N_c}=&\left[{6N_c\over (k+1) N_f}\right]^2-1~,\cr
\lambda_2'^*(k+1)=&2N_c^2\left[\left({6N_c\over (k+1) N_f}\right)
^2-1\right]-(N_c^2-1)\left[9\left({k-1\over k+1}\right)^2-1
\right]~.\cr
}}
Using the results of \refs{\IntriligatorJJ,\KutasovIY} one 
can show that \lamgenk\ leads to a sensible picture. 
In particular, the condition $\lambda_2'^*>0$ 
(which also implies $\lambda_2^*>0$) is the same as the 
condition that the ${\rm tr} X^{k+1}$ perturbation \xkpone\ 
is relevant in adjoint SQCD with vanishing superpotential\foot{The 
latter condition is given by eq. (3.15) in \KutasovIY\ for large
$N_f$, $N_c$.}. We see that the fixed point \valrxq\ exists if
and only if the operator \xkpone\ corresponds to a relevant 
perturbation of the IR limit of adjoint SQCD with $W=0$, something
that, as mentioned above, is not apriori obvious.

\newsec{Generalizations}

Many of the examples of RG flows in which the central charge $a$ was 
observed to decrease in \refs{\IntriligatorJJ-\IntriligatorMI}
fall into the class of models satisfying the assumptions of sections
2, 3 of this paper, and thus the behavior of $a$ for them is explained
by our analysis. However, there are other cases studied in 
\refs{\IntriligatorJJ-\IntriligatorMI} which do not satisfy some of the
assumptions on which our proof was based, and it is natural to ask
what happens in these cases. We will not attempt to provide a general
understanding of these issues here, but just comment on some of them.

The assumption that the IR $U(1)_R$ of an RG flow is a symmetry
of the full theory is certainly one that we would like to relax. 
Our techniques are actually useful for treating
cases in which it is violated. We next describe an example which,
as discussed in \KutasovIY, provides a sensitive test of the
$a$-theorem. 

In the last section we outlined the structure of
adjoint SQCD in the presence of a superpotential \xkpone\
that goes like ${\rm tr}X^{k+1}$. In particular, we saw that
despite appearances, this deformation can be relevant and
drive the system to a non-trivial fixed point. 

Suppose we now further deform this model by adding to
the superpotential a term that goes like ${\rm tr}X^k$, and
study the superpotential
\eqn\wkmone{W=g_k {\rm tr} X^{k+1}+g_{k-1}{\rm tr} X^k~.}
The UV fixed point of the RG flow we are interested in 
is the IR fixed point to which the system is driven by
the ${\rm tr} X^{k+1}$ superpotential \xkpone. At this 
fixed point, the operator $g_{k-1}{\rm tr} X^k$ is relevant, 
and adding it to the superpotential \wkmone\ leads to a new 
fixed point, in which the superpotential is effectively 
\eqn\irsup{W_{IR}\simeq g_{k-1}{\rm tr} X^k~,}
and the R-charge of $X$ is $R_X^{(IR)}=2/k$, 
in contrast to the UV fixed point \xkpone, where 
$W=g_k {\rm tr} X^{k+1}$ and $R_X=2/(k+1)$.

A natural question is whether $a$ decreases along this
RG flow. As pointed out in \AnselmiYS, this is far from
obvious. It was checked to be true in \KutasovIY\ by using
the results of \IntriligatorJJ. Here we would like to see
whether the fact that $a$ decreases in this flow can be understood
more conceptually, from the general perspective of sections 2, 3.

The technical complication here is that the IR R-symmetry, for
which $R_X=2/k$, is not a symmetry of the full theory -- it is
broken by the $X^{k+1}$ term in \wkmone. It is rather an accidental
symmetry of the IR theory, associated with the fact that $g_k\to 0$
in the IR. Nevertheless, the construction of sections 2, 3 is useful
for showing that $a$ decreases along this RG flow. To see that one
can proceed as follows. 

As in eq. \genalamn, the generalized central charge for the system
with superpotential \wkmone\ has the form 
\eqn\genkmone{\eqalign{
&a(R_X,R_Q,\lambda,\lambda_1,\lambda_2)=2(N_c^2-1)+\cr
&(N_c^2-1)\left[3(R_X-1)^3-(R_x-1)\right]+
2N_fN_c\left[3(R_Q-1)^3-(R_Q-1)\right]-\cr
&\lambda\left[N_cR_X+N_f(R_Q-1)\right]-
\lambda_1\left[2-(k+1)R_X\right]-\lambda_2\left[2-kR_X\right]\cr
}}
Maximizing with respect to $R_X$, $R_Q$ gives 
\eqn\rrrch{\eqalign{
&R_X(\lambda,\lambda_1,\lambda_2)=1-{1\over3}
\left(1+{\lambda N_c-\lambda_1(k+1)-\lambda_2k
\over N_c^2-1}\right)^{1\over2}\cr
&R_Q(\lambda,\lambda_1,\lambda_2)=1-{1\over3}\left(1+{\lambda\over 2N_c}
\right)^{1\over2}\cr
}}
Substituting in \genkmone, one finds the central charge $a$ as
a function of $(\lambda,\lambda_1,\lambda_2)$ which, as discussed
above, provide a particular parametrization of the space labeled by
the couplings $(\alpha,g_k, g_{k-1})$.

The UV fixed point of the flow we are interested in,
corresponding to the superpotential \xkpone\ is at
\eqn\uvk{{\partial a\over\partial\lambda}=
{\partial a\over\partial\lambda_1}=0;\;\;\;\lambda_2=0~.}
The IR fixed point corresponding to $W=g_{k-1}{\rm tr} X^k$
is at
\eqn\irkmo{{\partial a\over\partial\lambda}=
{\partial a\over\partial\lambda_2}=0;\;\;\;\lambda_1=0~.}
We would like to show that the value of $a$ at the solution
of \uvk\ is larger than at the solution of \irkmo. 

As a first step, one can reduce $a$ to a function of two variables,
$\lambda_1$ and $\lambda_2$, by imposing the relation 
\eqn\solvelam{
{\partial a(\lambda,\lambda_1,\lambda_2)\over\partial\lambda}=0~,}
which is common to \uvk\ and \irkmo,
and using it to solve for $\lambda$ as a function of $\lambda_1$,
$\lambda_2$. Substituting $\lambda(\lambda_1,\lambda_2)$ into
$a$ we find a central charge $a(\lambda_1,\lambda_2)$, which satisfies
\eqn\alonetwo{\eqalign{
{\partial a\over\partial\lambda_1}=&
-\left[2-(k+1)R_X(\lambda_1,\lambda_2)\right]\cr
{\partial a\over\partial\lambda_2}=&
-\left[2-kR_X(\lambda_1,\lambda_2)\right]\cr
}}
The UV fixed point of our flow corresponds to 
$\partial_{\lambda_1}a=0$, $\lambda_2=0$; the
IR fixed point to 
$\partial_{\lambda_2}a=0$, $\lambda_1=0$.

Now, consider the behavior of $R_X(\lambda_1,\lambda_2)$
as a function of $(\lambda_1,\lambda_2)$. The point
$(\lambda_1,\lambda_2)=(0,0)$ corresponds to the IR
fixed point of adjoint SQCD with vanishing superpotential
studied in section 2. We have assumed above that the
perturbation \xkpone\ is relevant there:
\eqn\rxoo{R_X(0,0)<{2\over k+1}~.}
Now, the UV fixed point of our RG flow has $R_X=2/(k+1)$
(see \uvk, \alonetwo), while the IR fixed point has $R_X=2/k$. 

Consider the curve 
\eqn\curvekpo{R_X(\lambda_1,\lambda_2)={2\over k+1}}
in the $(\lambda_1,\lambda_2)$ plane.
It must intersect the $\lambda_1$ and $\lambda_2$ axes
at two points, $(\lambda_1^*,0)$ and $(0,\lambda_2^*)$,
respectively. As before,  $(\lambda_1^*,\lambda_2^*)$ are
positive. Similarly, the curve
\eqn\newcurvekk{R_X(\lambda_1,\lambda_2)={2\over k}}
intersects the axes at the points 
$(\lambda_1'^*,0)$ and $(0,\lambda_2'^*)$. Since at 
$(\lambda_1,\lambda_2)=(0,0)$, $R_X<2/(k+1)$, by continuity
one has
\eqn\lamineq{\lambda_1'^*>\lambda_1^*,\;\;\;
\lambda_2'^*>\lambda_2^*~.}
We would like to show that 
\eqn\ineqaaa{a(\lambda_1^*,0)>a(0,\lambda_2'^*)~.}
To show that, consider the following trajectory
connecting the two points. First go from 
$(\lambda_1^*,0)\to(0,\lambda_2^*)$ following the curve
\curvekpo. Along this trajectory, $\partial_{\lambda_1}a=0$,
and therefore $a$ decreases:
\eqn\adecreases{{da(\lambda_1(\lambda_2),\lambda_2)\over 
d\lambda_2}={\partial a(\lambda_1(\lambda_2),\lambda_2)\over
\partial\lambda_2}=-{2\over k+1}<0~,}
where in the last step we used the second line of \alonetwo.

Now proceed on the $\lambda_2$ axis, from 
$(0,\lambda_2^*)\to (0,\lambda_2'^*)$. 
Here, too, $a$ decreases, using the second line of \alonetwo\ and the
fact that $R_X<2/k$ for all $\lambda_2^*<\lambda_2<\lambda_2'^*$.
Thus, we conclude that the central charge $a$ decreases along the flow
from $W\sim {\rm tr}X^{k+1}$ to $W\sim {\rm tr}X^k$, as found in
\KutasovIY.

We see that the generalized central charge $a$ away from fixed points
is useful in cases that do not satisfy the assumptions used in sections 2, 3.
What replaced those assumptions in the example discussed here is the dynamical
assumption that the RG flow $k\to k-1$ associated with the superpotential
\wkmone\ has the property that $g_{k-1}\to 0$ in the UV and
$g_k\to 0$ in the IR. It is in fact likely that by thinking 
of RG flow as rolling down in the landscape $a(\lambda_1,\lambda_2)$
one can derive this assumption. It is also possible that one can 
generalize the basic idea to a wide range of circumstances. 
We will not attempt to do that here.

It would be nice to generalize our construction to other situations
which fall outside the range of validity of the discussion of sections 2, 3.
These include the following:
\item{(1)} Often, the IR fixed point of an RG flow has additional accidental
symmetries of a different kind than those encountered in the example above.
One class of such accidental symmetries is associated with fields
becoming free in the IR. In fact, such symmetries are relevant in 
the adjoint SQCD example discussed at the end of section 3, in a 
certain range of parameters (see \KutasovIY). The effect of such
accidental symmetries on $a$-maximization is well understood
\refs{\KutasovIY,\IntriligatorMI} and it should not be difficult to
include them in our construction as well. We will leave this interesting problem
for future work.
\item{(2)} We did not discuss the effect of Higgsing, which is another
way of generating RG flows. In \refs{\KutasovIY,\IntriligatorMI}
it was found that the $a$-theorem seems to hold for such flows
as well. It would be interesting to generalize our discussion to this case.
\item{(3)} As we mentioned above, our approach is based on the NSVZ
$\beta$-function, which is known to break down rather abruptly at strong
coupling. In all known situations of this sort, one has a Seiberg dual 
description of the physics, and the $a$-theorem is known to be satisfied. 
In order to incorporate these situations in our framework one needs a 
better understanding of Seiberg duality than is presently available. This 
is an interesting direction for further research.

\vskip 1cm
\noindent{\bf Acknowledgments:}
I thank D. Sahakyan for discussions.
This work was supported in part by DOE grant
DE-FG02-90ER-40560.

\listrefs
\bye